\documentclass[twocolumn,showpacs,preprintnumbers,amsmath,amssymb]{revtex4}

\usepackage{epsfig}
\usepackage{graphicx}
\usepackage{dcolumn}
\usepackage{bm} 
\newcommand{\beq}{\begin{equation}}
\newcommand{\eeq}{\end{equation}}
\newcommand{\beqa}{\begin{eqnarray}}
\newcommand{\eeqa}{\end{eqnarray}}

\newcommand{\om}{\omega}

\def\jpb#1{{ J.\ Phys.\ B} {\bf#1}} 
\def\pra#1{{ Phys.\ Rev. A\/} {\bf#1}}

\def\prl#1{{ Phys.\ Rev.\ Lett.} {\bf#1}}

\begin{document}

\title{Elliptical Polarization and Probability of Double Ionization}

\author{Xu Wang}
\email{wangxu@pas.rochester.edu}
\author{J. H. Eberly}
\affiliation{ Rochester Theory Center and the Department of Physics 
\& Astronomy\\
University of Rochester, Rochester, New York 14627}


\date{\today}

\begin{abstract} 
The degree of elliptical polarization of intense short laser pulses is shown to be related to the timing of strong-field non-sequential double ionization. Higher ellipticity is predicted to force the initiation of double ionization into a narrower time window, and this ``pins" the ionizing field strength in an unexpected way, leading to the first experimentally testable formula for double ionization probability as a function of ellipticity.
\end{abstract}


\maketitle

\begin{figure}[b!]
\includegraphics[width=5cm,height=2cm]{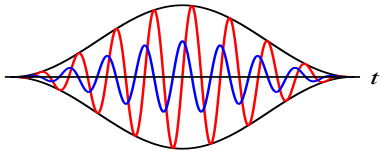}
\caption{{ \footnotesize \label{f.pulse} Illustration of an elliptically polarized laser pulse. The higher oscillating curve is the electric field along x and the lower curve is the electric field along y. The ellipticity is 0.5 here, and the full field is:  $\vec{E}(t) = E_0 f(t)[\hat{x}\sin(\om t + \phi) +  \varepsilon \hat{y} \cos(\om t + \phi)]$ for a smooth envelope $f(t)$ and random $\phi$.}}
\end{figure}

Atoms show an anomalously high degree of electron correlation in double ionization when exposed to femtosecond laser pulses in a wide range of intensities (10$^{14}$ - 10$^{16}$ W/cm$^2$) just below one atomic unit (see recent reviews \cite{Agostini-DiMauro, Becker-Rottke}). We are concerned here with effects of double ionization induced by elliptically polarized light (as illustrated in Fig. \ref{f.pulse}), over the entire range from linear to circular. 

There are two relevant double ionization channels: an atom may either lose two electrons one by one, which is called sequential double ionization (SDI), or lose the two electrons together in an e-e collision between a core electron and an already-ionized electron being driven back into the core by the reversed phase of the ionizing laser field, and this is called nonsequential double ionization (NSDI). 

It is considered highly unlikely or impossible for the second channel to produce any substantial degree of double ionization under elliptically polarized excitation. This was confirmed in early experiments  \cite{early polzn}. It is not hard to understand because a collision of the two electrons will be unlikely or impossible if the returning first-ionized electron is steered transversely off course by the ellipticity $\varepsilon$. 

However, this scenario is at odds with later experimental observations. Characteristic NSDI events have been observed under circular polarization with the molecules NO and O$_2$ \cite{Guo-Gibson} and with atomic magnesium \cite{Gillen-etal}. Elliptical polarization has also recently been predicted to have unexpected non-zero effects in SDI \cite{Wang-Eberly09PRL}, in agreement with experiment \cite{Maharjan-etal}. Evidently, polarization dependence has the potential for providing new insights into the complex character of two-electron correlation in double ionization \cite{Shvetsov-Shilovski-etal, Mauger-etal}.

\begin{figure}[h!]
\includegraphics[width=6cm,height=6cm]{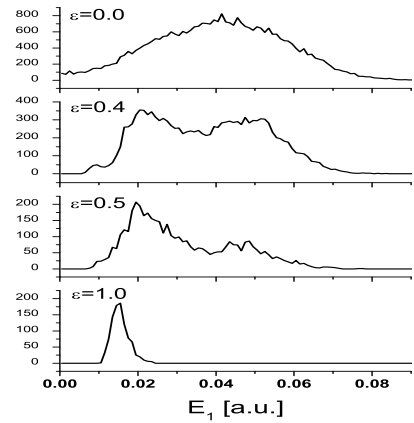}
\caption{{ \footnotesize \label{f.distrib-field} Distribution of field strengths at the time of first ionization for successful NSDI events, for several different laser field ellipticities.}}
\end{figure}

We report here ``experimental" evidence obtained for the dependence of NSDI on the degree of ellipticity of the incident pulse obtained via numerical simulations. Fig. \ref{f.distrib-field} shows one result, the distribution of initiating electric field strengths for successful NSDI events, for 4 values of $\varepsilon$. One sees in the top curve, obtained for linear polarization, that the electric field value at which first ionization occurs is widely distributed around a broad central peak. However, the data changes in a systematic way for larger values of the ellipticity. The distribution of ionizing fields splits into two peaks as $\varepsilon$ reaches 0.4 or 0.5 and becomes quite narrowly localized at less than half the distribution's previous peak value in the limit of circular polarization. 

We have obtained this data in a series of simulations that can be described in the same way as laboratory experiments. That is, our simulations represent a ``laboratory" in which a laser is focused into a target volume with one atom. The peak laser intensity is $6\times 10^{14}$ W/cm$^2$. After $10^7$ laser shots, assuming 100\% collection efficiency of ions created, there is data representing 10 million laser-atom exposures. From these, evidence for double ionizations must be extracted and analyzed in a systematic way. Since polarization dependence is of interest, an additional 10 million laser-atom exposures must be generated for every important value of ellipticity $\varepsilon$, say in 10 equal steps between 0 and 1. 

\begin{figure}[h!]
\includegraphics[width=6cm,height=6cm]{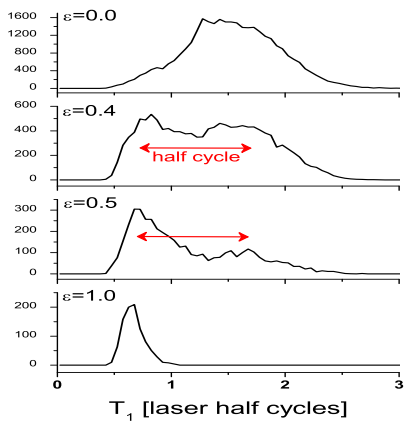}
\caption{{ \footnotesize \label{f.distrib-timing} Data showing first ionization times for successful NSDI events, for several different laser field ellipticities.}}
\end{figure}

The time at which the first ionization occurs is also widely distributed within the laser half-cycle at which the peak field reaches its ionizing value, as Fig. \ref{f.distrib-timing} shows. And the time of first ionization changes in a similar  fashion to the change in field strength, as $\varepsilon$ increases, first splitting and then narrowing and ending on a value almost exactly half a cycle earlier. This coordination of field strength behavior with the timing behavior is not hard to understand. The peak in timing should closely correspond to the peak in the distribution of field strengths, and if one is broad or narrow the other should be as well. In addition, the jump in timing by half a cycle is easily explained, since one expects NSDI events to originate near to field peaks, and these occur every half cycle.

However, the dependence of timing on ellipticity is unexpected and we believe not observed or previously suggested by theoretical considerations. It represents a variety of NSDI control not previously considered to be available. It is known that trajectories with different timing can play significantly different roles in high-field physical processes such as harmonic generation as well as double ionization. So far as we are aware, almost all previous investigations of NSDI timings have been restricted to linear polarization (an exception is the study by Shvetsov-Shilovsky, et al. \cite{Shvetsov-Shilovski-etal}). 

To begin to explain this dependence on ellipticity we make use of our demonstration \cite{Wang-Eberly10arX} that essentially every successful NSDI trajectory has transverse drift induced by the minor-axis component of polarization that is present, but which has been exactly compensated by a counter-acting transverse velocity present at initiation. Non-compensated trajectories are not able to recollide and are simply absent from any record of NSDI events. We believe that the same compensation mechanism is responsible for the ellipticity effect shown in Figs. \ref{f.distrib-field} and \ref{f.distrib-timing}.

For simplicity, as a first approximation we can assume that the  transverse momentum distribution available to the first electron at its instant of ionization is Gaussian: $P(v_y) \sim exp(-v_y^2/\Delta v_y^2)$. The compensation we referred to above is accomplished when one of these $v_y$ values matches the $\varepsilon$-promoted transverse drift velocity, which equals $\varepsilon E_1/\omega$, where $E_1$ is the field strength at ionization. Thus we will substitute $v_y = \varepsilon E_1/\omega$ in the Gaussian exponent. 

The overall probability of an NSDI event is then reasonably estimated as the product of the first electron's ``release probability" times the Gaussian probability of a velocity compensation:
\beqa \label{e.E1-epsilon}
P_{NSDI}(E_1, \varepsilon) &\sim& e^{-Q/E_1} \times e^{-(\varepsilon E_1/\omega)^2/\Delta v_y^2},
\eeqa
where the dependence of the first factor on ionization field strength is chosen in an obvious way, to mimic the main effect of $E_1$ in tunneling ionization, while $Q$ is a parameter related to ionization potential and varies by species. The same is probably true of $\Delta v_y^2$. 

\begin{figure}[b!]
\includegraphics[width=4cm]{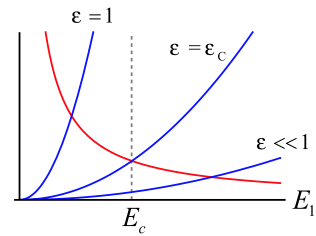}
\caption{{ \footnotesize \label{f.E1-epsilon} The two exponents in (\ref{e.E1-epsilon}) are separately plotted as functions of $E_1$. The second exponent is shown three times, for high, critical, and low values of ellipticity. The text explains the significance of the curve crossings.}}
\end{figure}

To examine in a na\"ive way the effect of $\varepsilon$ on the $E_1$ dependence of ionization we simply plot separately the two terms in the exponent in (\ref{e.E1-epsilon}) as a function of $E_1$. This is done in Fig. \ref{f.E1-epsilon}, where the second exponent contributes three curves for different values of $\varepsilon$. The maximum NSDI probability will come from the minimum value of the exponent, and from the graph in Fig. \ref{f.E1-epsilon} it is clear that this minimum occurs at, or very close to, the crossing of the curves. That is, a point substantially away from a crossing point will find either one or the other of the curves rising and making the exponent value a lot larger. An additional restriction is that $E_1$ should not exceed the critical over-the-barrier field $E_c$ because first ionization is expected to saturate under this field strength.  $E_c$ determines a critical ellipticity value $\varepsilon_c$: for $\varepsilon > \varepsilon_c$, the most probable $E_1$ value is determined by the crossing point; for $\varepsilon < \varepsilon_c$, the most probable $E_1$ value is simply $E_c$. 

However, the more important feature is the variation in the slopes of the curves at the crossing points. The slope is much smaller for the lowest values of $\varepsilon$, corresponding to near-linear polarization. Oppositely, the slopes are so high for crossings at or near to values of $\varepsilon \approx 1$ that the value of $\varepsilon$ practically pins the value of $E_1$. In other words, low ellipticities permit a wide range of $E_1$ values to create successful NSDI events, whereas near-circular ellipticities constrain $E_1$ to a very narrow range. This behavior is exactly what is needed to correspond to the dependence on $\varepsilon$ in Figs. \ref{f.distrib-field} and \ref{f.distrib-timing}. 

The significance of these findings goes beyond the explanation just discussed. We believe that they expose interesting unexplored territory within the NSDI domain by showing that new effects appear that depend on ellipticity. Prior to this, in the domain of extensive experimental activity, mostly confined to near-infrared wavelengths close to 800 nm, few theoretical studies of elliptically or circularly polarized pulses have been made (see \cite{Goreslavsky-Popruzhenko, Shvetsov-Shilovski-etal, Mauger-etal, Wang-Eberly10arX}).  

The simulation method that we used to obtain the ``experimental" data in Figs. \ref{f.distrib-field} and \ref{f.distrib-timing} is the classical ensemble method that has been described many times (see details in \cite{ClassicalEnsemble}), and its validity in interpretation of many double ionization phenomena has been presented \cite{Ho-etal05} and critiqued \cite{Rudenko-etal}. Classical modeling certainly misses true quantum features, but since the foundation for visualization of the NSDI channel is a classical view that attributes all action immediately after ionization to the classical force of the laser field on the freed electron as it returns to the vicinity of the ion core \cite{Kulander-etal, Corkum}, a classical model is not inappropriate for first analyses. 

In further support, one can say that among theoretical approaches the classical modeling used here is the most flexible and most widely applicable. For example, it is unique up to now among theoretical methods in finding agreement with prominently observed features of ion momentum spectra in triple ionization \cite{Zrost-etal, Ho-Eberly07} and with the first data on momentum spectra from double ionization under elliptical polarization \cite{Maharjan-etal, Wang-Eberly09PRL}. 

To summarize our use of the method quickly, a microcanonical ensemble of 10$^7$ members is generated using a many-pilot-atom method \cite{Abrines} before turning on the laser field. The energy of each 2e member of the ensemble is set to be -1.3 a.u., which is close to the binding energies of both Xe (-1.23 a.u.) and Kr (-1.41 a.u.) \cite{EnergyLevels}, and the wavelength is set at 780nm. The width of the familiar soft-Coulomb ``Rochester potential" \cite{ClassicalEnsemble} is taken as a = 1.77, which can be considered the model's single-parameter treatment of core effects that are species dependent (see also \cite{Haan-etal08}).  Experience with full-dimensioned calculations using this method \cite{Ho-Eberly06} has shown that out-of-plane effects can be neglected under current experimental conditions and we need only be concerned with the x-y plane, as is done here. 

\begin{figure}[t!]
\includegraphics[width=8cm]{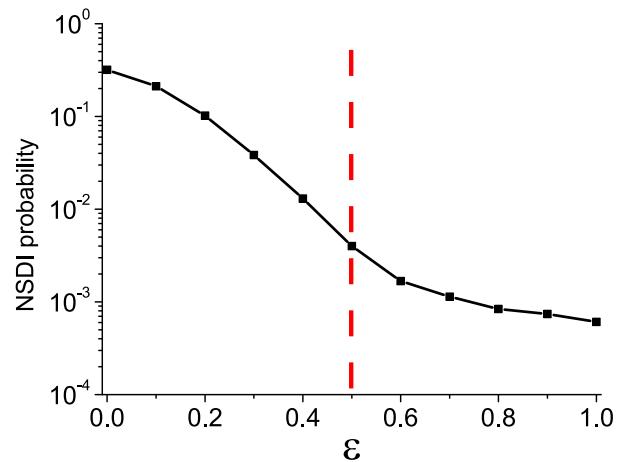}
\caption{{ \footnotesize \label{f.NSDIprob} NSDI probabilities for 10 values of ellipticity, from our numerical experiments. The connecting lines are only to guide the eye. An $\varepsilon_c$ value of about 0.5 is estimated from this figure, as indicated by the dashed line.}}
\end{figure}

Our large-ensemble simulations lead directly to NSDI probabilities for any value of $\varepsilon$, as shown in Fig. \ref{f.NSDIprob}. The results predict a dramatically slower decrease over the full range $0 < \varepsilon < 1$ for NSDI rates compared to those indicated by any previous theoretical considerations known to us. As Fig. \ref{f.NSDIprob} shows, they fall only 3 orders of magnitude and  remarkable slowly for the higher values of $\varepsilon$. We believe this provides the first explicitly $\varepsilon$-dependent explanation for the high NSDI rates in the high-ellipticity data from Guo and Gibson and Gillen, et al., mentioned already \cite{Guo-Gibson, Gillen-etal}. 

Our analysis of ellipticity dependence in NSDI production should be directly testable experimentally. To see this we exploit the ``pinning" of $E_1$ values mentioned above to eliminate $E_1$ in favor of $\varepsilon$ in the exponent of (\ref{e.E1-epsilon}). This provides a simple, even simplistic, formula for NSDI probability as a function of ellipticity. A quick check shows that for values of $E_1$ that are strongly pinned, for $\varepsilon > \varepsilon_c$ (the value of which can be estimated to be about 0.5 from Fig. \ref{f.NSDIprob}), our expression predicts that the exponent obeys a power law. Given expression (\ref{e.E1-epsilon}) as written, the power is +2/3, and the consequent distribution $exp(-\beta \varepsilon^{2/3})$ is fit in the high-$\varepsilon$ tail by a $\beta$ value in the neighborhood of 4 a.u.

In conclusion, we have shown that the degree of elliptical polarization of intense short laser pulses is related to the timing of strong-field non-sequential double ionization. Higher ellipticity is found to force the initiation of double ionization into a narrower time window, and this in turn ``pins" the ionizing field strength in an unexpected way. Among the consequences is an experimentally testable formula for double ionization probability as a function of ellipticity, which predicts a remarkably slow decrease in probability at high ellipticities, and we believe answers for the first time the standing need for an integrated theoretical explanation of the high NSDI rates under highly elliptical polarization in well-cited experiments \cite{Guo-Gibson,Gillen-etal}. The relationships discovered and reported here are generic, and were not tailored to a specific atom. But, as we have already mentioned, our classical modeling has been adequate in earlier studies for good semi-quantitative correspondence with multi-species NSDI effects, as recorded under linear polarization, and we expect the same will be true for non-zero ellipticity.

Acknowledgement: We appreciate a helpful communication from Prof. C. Guo. Partial financial support was provided by DOE Grant DE-FG02-05ER15713.


\begin{thebibliography}{99}
\bibitem{EnergyLevels} The binding energy of a 2e system is calculated as the negative sum of the 1st and 2nd ionization potentials. For example, the 1st and 2nd ionization potentials of Kr are 13.999 eV and 24.359 eV, which yields a binding energy of -(13.999+24.359) eV = -38.358 eV = -1.41 a.u.
\end{thebibliography}
\end{document}